\documentclass[conference]{IEEEtran}
\IEEEoverridecommandlockouts
\usepackage{cite}
\usepackage{amsmath,amssymb,amsfonts}
\usepackage{algorithmic}
\usepackage{graphicx}
\usepackage{textcomp}
\usepackage{xcolor}

\usepackage{graphicx}  
\usepackage{amsmath, amsopn, psfrag, amsthm} 
\usepackage{amssymb}
\usepackage{color}
\usepackage{psfrag}
\usepackage{epsfig}
\usepackage{breakcites}
\usepackage{amsthm}
\usepackage{lipsum}

\usepackage{cite}
\usepackage{amsmath,amssymb,amsfonts,amsthm}
\usepackage{algorithmic}
\usepackage{graphicx}
\usepackage{textcomp}
\usepackage{cite} 
\graphicspath{{./img/}}
\DeclareGraphicsExtensions{.pdf,.jpeg,.png}
\newtheorem{lemma}{Lemma}
\newtheorem{remark}{Remark}
\newtheorem{prop}{Proposition}

\def\BibTeX{{\rm B\kern-.05em{\sc i\kern-.025em b}\kern-.08em
    T\kern-.1667em\lower.7ex\hbox{E}\kern-.125emX}}
\begin{document}

\title{Composite IG/FTR Channel Performance in Wireless Communication Systems \\
\thanks{This work was funded in part by Junta de Andalucía, the
European Union and the European Fund for Regional Development FEDER
through grant P18-RT-3175, in part by "Consejería de Transformación
Económica, Industria, Conocimiento y Universidades" within the
talent acquisition program EMERGIA (ref. EMERGIA20\_00297), and in
part by MCIN/AEI/10.13039/501100011033 through grant
PID2020-118139RB-I00.}
\thanks{This work has been submitted to the IEEE for publication. Copyright may
be transferred without notice, after which this version may no longer be
accessible}
}

 \author{\IEEEauthorblockN{Maryam Olyaee\IEEEauthorrefmark{1}, Juan M.
Romero-Jerez\IEEEauthorrefmark{1}, F.
Javier~L\'opez-Mart\'inez\IEEEauthorrefmark{1}\IEEEauthorrefmark{2} and Andrea
J. Goldsmith\IEEEauthorrefmark{3}}
    \IEEEauthorblockA{\IEEEauthorrefmark{1}ComSP Lab, Telecommunication Research
Institute (TELMA), Universidad de M\'alaga, M\'alaga, 29010, (Spain)\\
    \IEEEauthorrefmark{2} Department of Signal Theory, Networking and
Communications, Universidad de Granada, 18071, Granada (Spain)\\
    \IEEEauthorrefmark{3}Dept. Electrical and Computer Engineering, Princeton
University, Princeton, NJ, 08544, USA\\} Email: maryam.olyaee@ic.uma.es,
romero@dte.uma.es, fjlm@ugr.es, goldsmith@princeton.edu.
    }

\maketitle

\begin{abstract}
We present a composite wireless fading model encompassing multipath fading and shadowing based on 
fluctuating two-ray (FTR) fading and inverse gamma (IG) shadowing. We first determine an 
alternative framework for the statistical characterization and performance evaluation of the FTR fading model, which is based 
on the fact that the FTR fading distribution can be described as an underlying Rician Shadowed (RS) distribution with 
continuously varying parameter $K_r$ (ratio of specular to diffuse components).
We demonstrate that this new formulation permits to obtain a closed-form expression of the generalized moment generating function (GMGF) of the FTR
model, from which the PDF and CDF of the composite IG/FTR model can be obtained in closed-form.
The exact and asymptotic outage probability of the IG/FTR model are analyzed and verified by Monte Carlo simulations.
	
\end{abstract}


\section{Introduction}
The random fluctuations affecting the radio signals have been classically divided into fast fading, resulting from multipath propagation,
and (slow) shadowing, caused by the presence of large objects like trees or buildings. These two effects are often characterized independently, 
however, fast fading and shadowing happen simultaneously at different time scales. Thus,
for a comprehensive analysis of wireless communications systems, a joint analysis is a more convenient approach by means of composite (i.e. integrated shadowing/fading) wireless 
channel models.

Regarding multipath fading, the Fluctuating Two-Ray (FTR) fading model was introduced in \cite{ftr} as a general model to capture the rapid fluctuations of
the radio channel in a wide variety of environments, through only three physically-motivated fading model parameters: $K$, $\Delta$ and $m$.
%
%
The FTR model provides a better
match to field measurements than most existing stochastic fading models in 
different environments, in particular at 28 GHz \cite{h18}. Consequently, the FTR fading model has been widely used for the 
performance analysis of 
millimeter wave (mmWave) communications  ~\cite{mine2,hadi1}.

Two different 
expressions for the probability density function (PDF)
of the signal-to-noise ratio (SNR) in FTR fading are available in the literature.
The original formulation for the PDF of the SNR of this model was introduced in \cite{ftr}, and an analytical expression based 
on the hypergeometric function $\Phi_2$ \cite{phi2} was given for the case of integer $m$. 
Later, an alternative expression for the PDF of the SNR valid for arbitrary $m$ was proposed in \cite{newR,correct} based on 
the use of infinite series, and therefore all the performance results based on this formulation entail evaluating an infinite series, which
must be truncated to be numerically computed, yielding error bounds, which in general will depend on the channel parameters values.

In this work we introduce an alternative framework for the FTR model that 
avoids the use of infinite series and is valid for arbitrary real $m$. The new formulation is based on the fact that
the distribution of the  FTR fading can be readily obtained
	by computing a finite-range integral over the distribution of the simpler Rician Shadowed (RS) fading model.
We demonstrate that this new framework permits to obtain a closed-form expression of the generalized 
	moment generating function (GMGF) of the FTR fading (which is a novel contribution of this work) that allows to readily obtain many different performance metrics 
	including channel capacity, symbol error rate,  outage probability, energy detection probability, secrecy outage probability, etc.


Moreover,  we use the derived GMGF expression to represent a composite channel model based on
 FTR fading and inverse gamma (IG) shadowing. We use the 
general approach recently introduced in \cite{Pablo}, where it was shown, through an extensive empirical validation based on data measurements
from a wide variety of scenarios, that the use of the IG distribution to model shadowing is well-justified in cases of
mild and moderate shadowing conditions, even providing a better match to empirical measurements than the more common log-normal distribution in some cases.
We will provide closed-form expressions for the PDF and CDF of the composite IG/FTR channel model, and calculate the outage probability, both in
exact and asymptotic forms.

The rest of this paper is organized as follows: In Section~II, the preliminary definitions are introduced.
Section~III presents the connection between the FTR and RS fading models.
	Based on this connection, a mathematical framework to characterize the outage probability analysis on IG/FTR composite model is shown in Sectionv IV. 
Finally, Section~V provides numerical and simulation results followed by the conclusions in Section~VI.

\section{Preliminary Definitions}
In the FTR fading model, the wireless channel consists of two fluctuating dominant waves, referred to as specular components, 
to which other diffusely propagating waves are added. The complex baseband received signal can be expressed as 
\begin{align}
	V = \sqrt{\zeta} V_1 \exp (j \phi_1) + \sqrt{\zeta} V_2 \exp (j \phi_2) +X+jY,\label{Vr}
\end{align}
where $ V_n $ and $\phi_n$, for $n=1,2$, represent, respectively, the average amplitude and
the uniformly distributed random phase of the n-\emph{th} specular component, such that $\phi _n \sim \mathcal{U}[0,2\pi)$. 
The term $X + jY$ is a complex Gaussian random variable, with $X,Y \sim \mathcal{N}(0,\sigma^2)$, representing the diffuse 
received signal component 
due to the combined reception of numerous weak scattered waves.
On the other hand, $\zeta$ is a unit-mean Gamma distributed random variable modulating the specular components, whose PDF is 
given by
\begin{align}
	f_{\zeta}(u) =\frac{m^m u^{m-1}}{\Gamma(m)} e ^{-mu}.
\end{align}
This model is conveniently expressed in terms of the parameters $K$ and $\Delta$, defined as
\begin{align}
	\Delta =\frac{2V_1 V_2}{V_1^2+V_2^2},\;\;
	K= \frac{V_1^2 +V_2^2}{2\sigma^2},
\end{align}
where $K$ denotes the power ratio of the specular components to the diffuse components, and 
$\Delta$ provides a measure of the similarity of the two specular components, ranging from 0 (one specular component is absent)
to 1 (the specular components have the same amplitude).
Additionally, we define the ancillary random variable $\theta \triangleq \phi_1-\phi_2$. Note that, as the phase difference is 
modulo $2\pi$, $\theta$ will be uniformly distributed, and therefore we can write $\theta \sim 
\mathcal{U}[0,2\pi)$.

Throughout this paper, we characterize the distribution of the instantaneous envelope power $\gamma={\left|V\right|^2}$ at the receiver or,
equivalently, the SNR with normalized noise power.

	\textbf{Definition 1:}
	A random variable $\gamma$ following a FTR distribution with parameters $m$, $K$, $\Delta$ and mean $\overline\gamma$ will be denoted by 
	$\gamma \sim \mathcal{FTR}(\overline{\gamma},m,K,\Delta)$, and its PDF will be denoted by $f_{\gamma}^{\rm FTR}(x;\overline{\gamma},m,K,\Delta)$, where the 
	parameters may be dropped from the notation when there is no confusion.
	
	\textbf{Definition 2:}
	A random variable $\gamma$ following a squared RS distribution with parameters $m$, $K_r$ and $\overline{\gamma}$ will be  denoted  by 
	$\gamma \sim \mathcal{RS} (\overline{\gamma},m,K_r)$, and its PDF can be written as
	\begin{align}\nonumber \label{fRS}
		&f_\gamma ^{\rm RS}(x;\overline \gamma, m,{K_r}) = {\left( {\frac{m}{{m + {K_r}}}} \right)^m}\frac{{1 + {K_r}}}{{\overline \gamma }} \\
		&\times  \exp \left( { - \frac{{1 + {K_r}}}{{\overline \gamma }}x} \right) {_1}{F_1}\left( {m;1;\frac{{{K_r}\left( {1 + {K_r}} \right)}}{{\overline \gamma \left( {m + {K_r}} \right)}}x} \right),
	\end{align}
	where $_1 F_1(\cdot)$ is the confluent hypergeometric function of the first kind \cite[eq. (9.210.1)]{Gradsh}, $\overline{\gamma}=2\sigma^2 (1+K_r)$, and $K_r=\frac{\Omega}{2\sigma^2}$,  where $\Omega$ and $2\sigma ^2 $ are the powers of the specular and diffuse components, respectively.

\section{FTR Formulation as a Continuous Mixture of RS Variates}

In this section, we explore the connection between the FTR and the RS fading models. We demonstrate
that the statistics of the RS fading can readily be extended to the FTR case.

\subsection{Connection between the FTR and the RS fading models}
The PDF of the power of a wireless channel consisting of $N$ specular components and a diffuse 
complex Gaussian component was obtained in \cite{Romero19} in  terms of the powers of the individual components. 
Thus, for the case $N=2$, the PDF of a random variable $\gamma \sim \mathcal{FTR}(\overline{\gamma};m,K,\Delta)$, 
can be written as continuous mixture of squared RS variates as 
\begin{align}
	f_\gamma ^{\rm FTR}(x;\overline{\gamma},&m,K,\Delta ) \nonumber \\& 
	= \frac{1}{\pi}\int_0^{\pi } {f_{\gamma | \theta}^{\rm RS}(x;\overline{\gamma},m,K\left( {1 + \Delta \cos (\theta )} \right)d\theta }, \label{RS2FTR}
\end{align}
with $\gamma | \theta \sim \mathcal{RS} (\overline{\gamma},m,K\left( {1 + \Delta \cos (\theta )} \right))$.
\begin{remark} \label{rm1}
It must be noted that the factor $\frac{{\overline \gamma (\theta)}}{{1 + K_r(\theta)}}$  remains invariant in the transformation defined in 
\eqref{RS2FTR}, and  in all subsequent expressions derived from it. This is due to the fact that 
parameters $K$ and $	\overline \gamma $ of the FTR model are related by the expression
$	\overline \gamma  = \left( {V_1^2 + V_2^2 + 2{\sigma ^2}} \right) = 2{
\sigma ^2}\left( {1 + K} \right)$. On the other hand, parameter $K_r$ of the RS model in \eqref{RS2FTR}
is a function of the particular realization of the RS fading, as it verifies 
 $K_r(\theta)=K\left( {1 + \Delta \cos (\theta )} \right)$. Therefore, the average power for the RS fading will depend also on $\theta$
according to 
$
	\overline \gamma (\theta ) = 2{\sigma ^2}\left( {1 + K_r (\theta )} \right),
$
yielding,
\begin{align}
{\rm{ }}\frac{{\overline \gamma (\theta )}}{{1 + K_r(\theta )}} = 2{\sigma ^2} 
=\frac{{\overline \gamma }}{{1 + K}} . \label{eqg}
	\end{align}
\end{remark}

\begin{prop} \label{propRSFTR} 
	Let $\gamma \sim \mathcal{FTR}(\overline{\gamma},m,K,\Delta)$, then, its PDF can be computed as 
	\begin{align}
		\nonumber
		&	 f_\gamma ^{\rm FTR}(x;\overline{\gamma},m,K,\Delta )
		\\ \nonumber& =\frac{1}{\pi} \int_0^{\pi } {{\left( {\frac{m}{{m + K\left( {1 + \Delta \cos (\theta )} \right)}}} \right)}^m} \exp \left( { - \frac{{1 + K}}{{\overline \gamma }}x} \right)  \nonumber\\ &
		\times \frac{{1 + K}}{{\overline \gamma }}	{_1}{F_1}\left( {m;1;\frac{{\left( {1 + K} \right)K\left( {1 + \Delta \cos (\theta )} \right)}}{{\overline \gamma m + \overline \gamma K\left( {1 + \Delta \cos (\theta )} \right)}}x} \right)d\theta. \label{fFTR}
	\end{align}
\end{prop}
\begin{proof}
	 This expression is obtained by plugging \eqref{fRS} into (\ref{RS2FTR}) and considering \eqref{eqg}. 
\end{proof}
It was demonstrated in \cite{ftr} that the FTR fading model collapses to the Hoyt fading model for $m=1$. In this regard,
it can be easily shown that the integral connection
between the Hoyt and Rayleigh models presented in \cite{NewF} is actually a particular case of \eqref{fFTR} when $m=1$.
\begin{remark} \label{rm2}
	Proposition \ref{propRSFTR} implies that, conditioning on $\theta$, the FTR distribution is actually a RS distribution, and we can write
	the conditional PDF as
	\begin{align}\nonumber
		&	f_{\gamma | \theta }^{\rm FTR}(x;m,K,\Delta ) = f_{\gamma | \theta}^{\rm RS}(x;m,K(1+\Delta cos (\theta)) ) =
		\\ \nonumber &
		{\left( {\frac{m}{{m + K\left( {1 + \Delta \cos (\theta )} \right)}}} \right)^m}\frac{{1 + K}}{{\overline \gamma }}
		\exp \left( { - \frac{{1 + K}}{{\overline \gamma }}x} \right)
		\\& \times  \;{_1}{F_1}\left( {m;1;\frac{{K\left( {1 + K} \right)\left( {1 + \Delta \cos (\theta )} \right)}}{{\overline \gamma m +
					K\overline \gamma \left( {1 + \Delta \cos (\theta )} \right)}}x} \right)\label{fFTRc},
	\end{align}
	which will be used in subsequent derivations.
\end{remark}

\begin{prop} \label{lemcdf} {
		Let $\gamma \sim \mathcal{FTR}(\overline{\gamma},m,K,\Delta)$, then, its cumulative distribution function (CDF) can be calculated as }
	\begin{align}\nonumber
		&	F_{\gamma}^{\rm FTR}(x;m,K,\Delta ) \\& \nonumber
		=\frac{1}{\pi} \int_0^{\pi} \frac{{1 + K}}{{\overline \gamma }}x{\left( {\frac{m}{{m + K\left( {1 + \Delta \cos (\theta )} \right)}}} \right)^m} \times
		\\&
		{\Phi _2}\left( {1 - m, m;2; - \frac{{1 + K}}{{\overline \gamma }}x; \frac{{-\left( {1 + K} \right)mx}}{{\overline \gamma (m + K \left( {1 + \Delta \cos (\theta )} \right))}}} \right) d\theta,
		\label{CDF}
	\end{align}
	where $\Phi_2$ is the bivariate confluent hypergeometric function defined in \cite[p. 34, eq. (8)]{phi2}. 
	{An efficient algorithm to compute this function can be found in \cite{Matlabprog}.}
\end{prop}
\begin{proof}
	{ This expression is readily obtained from \cite{Romero19} for the case $N=2$ and considering \eqref{eqg}.}
\end{proof}

\subsection{Generalized MGF}
The generalized MGF is an important statistical function relevant to wireless communication theory, as it 
naturally appears when analyzing different scenarios such as energy detection probability, outage probability with co-channel 
interference in interference limited scenarios, physical layer 
security analysis \cite{beckmann}, or in the context of composite fading channel modeling \cite{Pablo}.
The GMGF of a random variable $X$ is defined: 
\begin{align}
	&{M^{(n)}_X}(s) = \int_0^\infty  {{x^n}\exp \left( {xs} \right)f_X(x)dx}\label{gmgfd}.
\end{align}
Note that in the case of $n\in \mathbb{N}$, the generalized MGF coincides with the $n^{th}$ order derivative 
of the MGF.
\begin{lemma}\label{lemGMGF}
	Let  $\gamma \sim \mathcal{FTR}(\overline{\gamma};m,K,\Delta)$, then, the GMGF of $\gamma$ can be obtained in closed-form
	as (\ref{gmgf}),
	\begin{figure*} [!t]
		\small
		\begin{align}\nonumber
			M_{\gamma}&^{(n)}(s)= n!{{{m^m}{{\left( {1 + K - \overline \gamma s} \right)}^{m-n-1}}}}
			{{\overline \gamma }^n}\sum\limits_{l = 0}^n {\binom{n}{l}\frac{{{{\left( m \right)}_l}}}{{l!}}} 
			\frac{{( {1 + K} )}^{l + 1}K^l}{{{{\left[ {m\left( {1 + K} \right) - \left( {m + K - K\Delta } \right)\overline \gamma s} 
							\right]}^{m+l}}}} \nonumber\\ 	& 
			\times
			\sum\limits_{q = 0}^l 
			\binom{l}{q}
			{
				\left( {1 - \Delta } \right)^{l - q}}
			{\left( {2\Delta } \right)^q}
			\frac{{\Gamma \left( {\frac{1}{2} + q} \right)}}{{\sqrt{\pi}\Gamma (q + 1)}} 
			{{\kern 1pt}_2}{F_1}\left( {{m+l},\frac{1}{2} + q;q + 1;\frac{{2K\Delta \overline \gamma s}}{{m\left( {1 + K} \right) - \left
						( {m + K - K\Delta } \right)\overline \gamma s}}} \right).\label{gmgf}
			\\
			\hline \nonumber
		\end{align}
	\end{figure*}
	where $_2F_1(\cdot)$ is the Gaussian hypergeometric function \cite[eq. (9.100)]{Gradsh} and where $(a)_n$ is the Pochhammer symbol.
\end{lemma}
\begin{proof}
The conditional generalized MGF 
 can be written by plugging (\ref{fFTRc}) into \eqref{gmgfd}, i.e.:
\begin{align}\nonumber
	{M^{(n)}_{\gamma|\theta}}(s) &= 
\int_0^\infty  {x^n}\exp \left( {xs} \right)
			{\left( {\frac{m}{{m + K\left( {1 + \Delta \cos (\theta )} \right)}}} \right)^m}
				\\& \nonumber
		\times	\frac{{1 + K}}{{\overline \gamma }}
	\exp \left( { - \frac{{1 + K}}{{\overline \gamma }}x} \right)
		\\&  \times \;{_1}{F_1}\left( {m;1;\frac{{K\left( {1 + K} \right)\left( {1 + \Delta \cos (\theta )} \right)}}{{\overline \gamma m + K\overline \gamma \left( {1 + \Delta \cos (\theta )} \right)}}x} \right)
				dx. \label{mgfc}
\end{align}
The integral in \eqref{mgfc} can be solved with the help of \cite[eq. (7.621.4)]{Gradsh} as
\begin{align}\nonumber
&	M_{\gamma \left| \theta  \right.}^{(n)}(s)= \\ \nonumber
&  {\left( {\frac{m}{{m + K + K\Delta \cos (\theta )}}} \right)^m}\frac{({1 + K})\Gamma (n + 1)}{{{{\left( {1 + K - \overline \gamma s} \right)}^{n + 1}}}}  
	{{\overline \gamma }^n} \times
	\\& \nonumber
{_2}{F_1}\left( {m,n + 1;1;\frac{{\left( {1 + K} \right)\left( {K + K\Delta \cos (\theta )} \right)}}{{\left( {m + K + K\Delta 
\cos (\theta )} \right)\left( {1 + K - \overline \gamma s} \right)}}} \right)\label{mmgfn}
\\ \nonumber
& = {\left( {\frac{m(1+K-\overline{\gamma}s)}{m(1+K)-({m + K + K\Delta \cos (\theta )})\overline{\gamma} s}} \right)^m}
\\&\times 
\frac{({1 + K})\Gamma (n + 1)}{{{{\left( {1 + K - \overline \gamma s} \right)}^{n + 1}}}}  
	{{\overline \gamma }^n} \times
\nonumber	\\ 
	& 
	 \;{_2}{F_1}\left( {m,-n;1;
		-\frac{{\left( {1 + K} \right)\left( {K + K\Delta \cos (\theta )} \right)}}{m(1+K)-({m + K + K\Delta \cos (\theta )})\overline{\gamma} s}} \right),
\end{align}
where the last equality is obtained with the help of  \cite [eq. (9.131)]{Gradsh}. 
Besides, based on 
\cite[p.17 eq. (12)]{phi2}
for integer $n$, the hypergeometric function can be written as
\begin{align} \nonumber
		_2{F_1}\left( { m,-n;c;z} \right)&=
	\;	_2{F_1}\left( { - n,m;c;z} \right) \\&= 
		\sum\limits_{l = 0}^n {{{( - 1)}^l}} \binom{n}{l}
		\frac{{{{\left( m \right)}_l}}}{{{{\left( c \right)}_l}}}{z^l},
\end{align}
thus, the conditional GMGF is determined as
	\begin{align}
		\nonumber
		M_{\gamma \left| \theta  \right.}^{(n)}(s)& = n!\frac{{{m^m}{{\left( {1 + K - \overline \gamma s} \right)}^m}}}{{{{\left( {1 + K - \overline \gamma s} \right)}^{n + 1}}}}{{\overline \gamma }^n}\sum\limits_{l = 0}^n {\binom{n}{l}\frac{{{{\left( m \right)}_l}}}{{l!}}} 
		\\&	\times
		\frac{{{{\left( {1 + K} \right)}^{l + 1}}K^l{{\left( {1 + \Delta \cos (\theta )} \right)}^l}}}{{{{\left[ {m\left( {1 + K} \right) - \left( {m + K + K\Delta \cos (\theta )} \right)\overline \gamma s} \right]}^{l + m}}}}.\label{gmgfC}
	\end{align}
	Therefore, the unconditional GMGF will be given by
	\begin{align}\nonumber
		M_{\gamma}^{(n)}	= \frac{1}{\pi} \int_0^{\pi}  M_{\gamma \left| \theta  \right.}^{(n)}(s) d \theta,
	\end{align}
	and with the help of integral I1 solved in the Appendix the proof is completed. 
\end{proof}

It should be noted that \eqref{gmgf} is valid for all possible values of the channel parameters when 
$s$ is a non-positive real number (which is usually the case in communication theory applications),
since for the Gaussian hypergeometric function $_2 F_1(a,b;c;z)$ in the expression it is always true that $z \in \mathbb{R}$ with $z \leq 0$,
which permits its computation using its integral representation \cite[eq. (9.111)]{Gradsh}.
The moments of the distribution can be easily obtained from the GMGF by simply setting $s=0$.

\section{ A Composite IG/FTR Fading Model}
We now use the obtained GMGF to represent the composite IG/FTR channel model using the general approach in \cite{Pablo}.

The received power of the composite IG/FTR model can be expressed as 
\begin{align}
	Z = \bar Z \mathcal{G}\mathcal{V},
\end{align}
where $\bar Z=\mathbb{E}\{Z\}$, and $\mathcal{G}$, $\mathcal{V}$
are independent random variables representing the shadowing and the fast fading with normalized power, respectively.
Thus, $\mathcal{G}$ is an IG random variable with shape parameter ${\lambda}$ and $\mathbb{E}\{\mathcal{G}\}=1$, and $\mathcal{V}$ is the FTR random variable with parameters $m,K,\Delta$, and $\bar\gamma=1$.

We will provide the PDF and CDF of the composite IG/FTR and obtain the outage probability. The PDF of $Z$ can be obtained  from the closed-form expression  of the GMGF in \eqref{gmgf} and \cite[eq. 12]{Pablo} as
\begin{align}
	f_Z(z) = \frac{\bar{Z}^{\lambda}(\lambda-1)^{\lambda} }{z^{\lambda+1}\Gamma(\lambda)} M_{\mathcal{V}}^{(\lambda)} \left( \frac
{(1-\lambda)\bar Z}{z}\right),\label{fz}
\end{align}
and the CDF of $Z$ for integer $\lambda$, by using \cite[eq. 22]{Pablo} and \eqref{gmgf}, is expressed as
\begin{align}
	F_Z(z) = \sum_{n=0}^{\lambda-1} \frac{\bar{Z}^{n}(\lambda-1)^{n} }{z^{n}\Gamma(n+1)} M_{\mathcal{V}}^{(n)}  \left(\frac{(1-
\lambda)\bar Z}{z}\right).\label{cdfz}
\end{align}

The PDF and CDF of the received signal amplitude $R$ can be obtained by means of the transformation $Z=R^2$, from which 
they can be calculated, respectively, as $f_R(r)=2rf_Z(r^2)$ and $F_R(r)=F_Z(r^2)$.

Considering that the SNR at the receiver is defined as
\begin{align}
	\gamma_z =\frac{ \bar{\gamma}_z Z}{ \bar Z},
\end{align}
	where $\bar{\gamma}_z =\mathbb E\{\gamma_z\}$, then the outage probability is given by \cite{gbook}
\begin{align}
	P^{\rm IG/FTR}_{\rm out} (\gamma_{th}) \triangleq P(\gamma_z<\gamma_{th})=F_{\gamma_z}(\gamma_{th})= F_Z(\bar Z \gamma_{th}/\bar \gamma_z),
\end{align}
where  $\gamma_{th}$ is the SNR threshold for reliable communication, and the CDF $F_Z(z)$ is given in \eqref{cdfz}.
Therefore, the asymptotic outage probability can be achieved by $\bar\gamma_z \to \infty$. Using an asymptotic approximation for the FTR CDF in \cite[eq. 18]{HyperFTR2019} and the asymptotic outage probability for IG-based composite fading models for $\lambda>1$ in \cite[eq. 33]{Pablo}, the composite IG/FTR asymptotic outage probability is given by
\begin{align}\nonumber
	P^{\rm IG/FTR}_{\rm out}& (\gamma_{th}|\bar\gamma_z \to \infty)\approx 
	\frac{\Gamma(\lambda+1)}{\Gamma(\lambda)(\lambda-1)}
	\frac{(1+K)}{(1+\frac{K}{m})^m}\;
	\\& \times _2F_1(\frac{m}{2},\frac{1+m}{2};1;\frac{\Delta^2}{(\frac{m}{K}+1)^2}) 
	\frac{\gamma_{th}}{\bar\gamma_z},\label{asym_z}
\end{align}
where $_2F_1(.)$ is the Gauss hypergeometric function.

\section{Numerical results}
In this section, numerical results obtained from our analytical derivations are presented for the new expression of
the PDF of the FTR model, as well as for the PDF and the outage probability of the composite IG/FTR wireless channel model
defined in the previous section.
Results are validated by Monte Carlo simulations, showing an excellent agreement. 

\begin{figure}[t]
	\centering
	\includegraphics[width=0.99\linewidth]{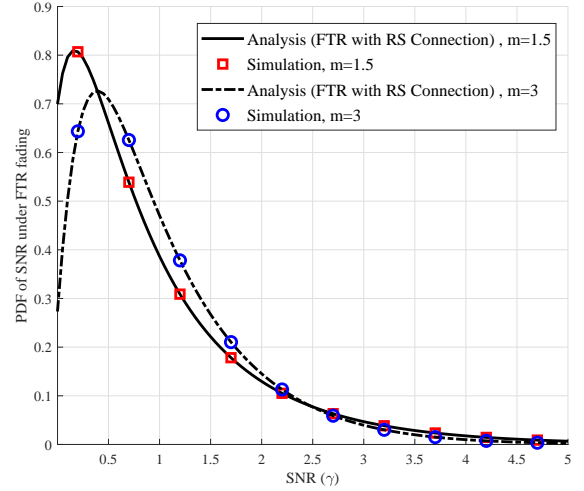}
	\caption{ Analysis and Monte-Carlo simulation of the PDF of $\gamma$ under FTR fading obtained using the FTR-RS connection for $m=1.5,3$.}
	\label{pdf}
\end{figure}

In Fig. \ref{pdf}, the PDF of the FTR fading model is plotted for two values of the $m$ parameter: one integer ($m=3$) and one  non-integer $m=1.5$ which has been obtain from \eqref{fFTR}. Simulation 
results show a perfect match to the analytical results in all cases.

\begin{figure}[t]
	\centering
	\includegraphics[width=0.95\linewidth]{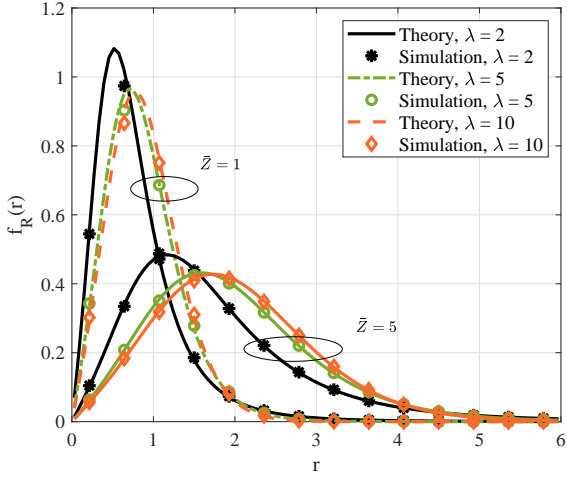}
	\caption{Analytical and simulation results for the signal amplitude PDF of composite IG/FTR model with parameters $\bar{\gamma}=1$, $K=4$, $\Delta = 0.2$, and $m = 2$.}
	\label{figfZ}
\end{figure}

 Fig.  \ref{figfZ} depicts the PDF of the received signal amplitude for the IG/FTR channel model, for different values of the 
gamma distribution parameter $\lambda$ and $\bar{Z}$. The theoretical plots  perfectly agreed with the Monte-Carlo simulation. 
Independently of the FTR model parameters, it can be observed that smaller values of $\lambda$ 
make PDFs more sparse, and larger values of $\lambda$ decrease the shadowing severity. This effect is shown for 
different parameters of $\bar Z =1,5$.

\begin{figure}[t]
	\centering
	\includegraphics[width=0.95\linewidth]{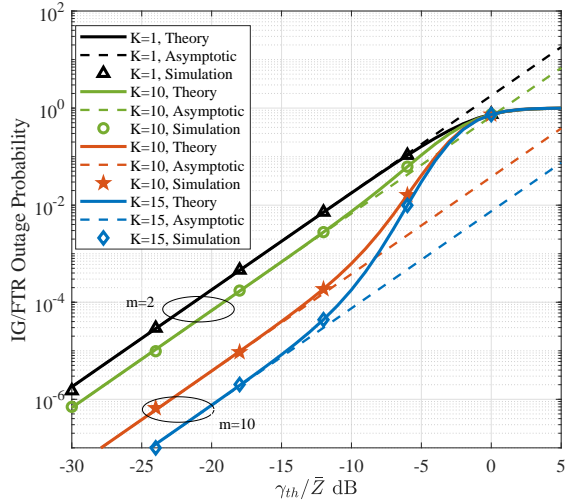}
	\caption{Analytical, asymptotic and simulation results for the outage probability of composite IG/FTR vs. $\gamma_{th}/\bar{Z}$ with parameters $\bar{\gamma}=1$, $\Delta = 0.3$, and $\lambda = 2$.}
	\label{figPout_z}
\end{figure}

Finally, Fig. \ref{figPout_z} plots the outage probability over IG/FTR fading in terms of the normalized threshold for
different values of the FTR parameters ($m$ and $K$). It can be observed that the asymptotic expression in \eqref{asym_z} 
perfectly fits the theoretical curves as $\gamma_{th}/\gamma_z \to 0$. As expected, increasing the $K$ parameter, representing 
the power ratio of the specular to diffuse components, decreases the outage probability. Also, the outage probability declines for 
light ($m = 10$) fluctuations in comparison to strong ($m = 2$) fluctuations of the specular components. On the other, the diversity order 
remains the same independently of the FTR parameters, as shown by the same slope in the high SNR regime (high values of $\bar Z$).

\section{Conclusion}
We presented a flexible connection that describes the relationship between the Fluctuating Two-Ray (FTR) and Rician Shadowed 
fading models. Based on this novel formulation, we provided for the first time a closed-form expression of the
GMGF for FTR fading. As an application of  these 
results, the chief statistical functions as well as the exact and asymptotic outage probability 
of a composite IG/FTR wireless channel model have been provided in closed-form.

\section*{Appendix 
	\\
	 Solving Integral I1 }

 { The integral I1 for positive integer $P1$, arbitrary positive $P2$, arbitrary $\alpha$ and $|\beta|<1$ can be computed as}
    \begin{align}
&  \text{I1} \triangleq \int_0^\pi  \frac{\left( {1 + \alpha \cos (\theta )} \right)^{P1}}{\left( {1 + \beta \cos (\theta )} \right)^{P2}}d \theta
  	\\ \nonumber
  &\overset{(A)}{=}
  \frac{1}{(1-\beta)^{P2}}\int_0^1 
  (1-\alpha+{2\alpha}x)^{P1} 	\left( 1+\frac{2\beta}{1-\beta}x \right)^{-P2}
 \\&  \times 
  (-1)x^{-\frac{1}{2}}(1-x)^{-\frac{1}{2}}
  dx \\&  
\overset{(B)}{=} 
  \frac{-1}{(1-\beta)^{P2}}
  \sum_{q=0}^{P1} \binom{P1}{q} ({2\alpha})^q (1-\alpha)^{P1-q}
\nonumber\\&  \times
  \int_0^1  
  x^{q-\frac{1}{2}}(1-x)^{-\frac{1}{2}} 	\left( 1+\frac{2\beta}{1-\beta}x \right)^{-P2}
  dx \\&  
\overset{(C)}{=} 
  \frac{-1}{(1-\beta)^{P2}}
  \sum_{q=0}^{P1} \binom{P1}{q} 
  ({2\alpha})^q (1-\alpha)^{P1-q}
  \frac{\sqrt{\pi}\Gamma\left(q+\frac{1}{2}\right)}{\Gamma(q+1)}
  \nonumber \\&  \times 
  {{\kern 1pt}_2}{F_1}\left(P2,q+\frac{1}{2} ;q + 1;-\frac{2\beta}{1-\beta} \right),
\end{align}
where $(A)$ followed from the change of variables $\cos (\theta )=2x-1$, yielding $d\theta = \frac{-dx}{\sqrt{x}\sqrt{1-x}}$, $(B)$ 
followed by using the binomial theorem $(a+b)^n=\sum_{q=0}^n \binom{n}{q}a^{n-q}b^q$ for positive integer $P1$, and $(C)$ is 
obtained from the integral representation of the Gauss hypergeometric function \cite[eq. (9.111)]{Gradsh}
\begin{align}
	&_2{F_1}\left( {a,b;c;z} \right)  \nonumber\\&
= \frac{{\Gamma (c)}}{{\Gamma (b)\Gamma (c - b)}}\int_0^1 {{t^{b - 1}}{{\left( {1 - t} \right)}^{c - b - 1}}{{\left( {1 - tz} 
\right)}^{ - a}}dt}  \label{hyper}
\end{align}
where $a=P2$, $b=q+\frac{1}{2}$, $c=q+1$ and $z=-\frac{2\beta}{1-\beta}$.

\end{document}